# Integrated astrophotonic phase control for high resolution optical interferometry

Ross Cheriton[a,*], Siegfried Janz[a], Glen Herriot[b], Jean-Pierre Véran[b], Brent Carlson[c]
[a]Advanced Electronics and Photonics, National Research Council of Canada, 1200 Montreal Rd, Building M-50, Ottawa, ON, Canada, K1A 0R6,
[b]Herzberg Astronomy and Astrophysics, National Research Council of Canada 5071 W. Saanich Rd, Victoria, BC, Canada, V9E 2E7
[c]Herzberg Astronomy and Astrophysics, National Research Council of Canada 717 White Lake Road, Kaleden, BC, Canada, V0H 1K0
*ross.cheriton@nrc-cnrc.gc.ca;

## ABSTRACT

Long baseline optical interferometry and aperture synthesis using ground-based telescopes can enable unprecedented angular resolution astronomy in the optical domain. However, atmospheric turbulence leads to large, dynamic phase errors between participating apertures that limit fringe visibility using telescopes arrays or subaperture configurations in a single large telescope. Diffraction limited optics or adaptive optics can be used to ensure coherence at each aperture, but correlating the phase between apertures requires high speed, high stroke phase correction and recombination that is extremely challenging and costly. As a solution, we show an alternative phase correction and beam combination method using a centimeter-scale silicon astrophotonic chip optimized for H-band operation.

The 4.7x10mm silicon photonic chip is fabricated using electron beam lithography with devices with 2 up to 32 independent channels. Light is coupled into the chip using single mode fiber ribbons. An array of microheaters is used to individually tune the effective index of each spiral delay waveguides. Narrowband spectral splitters at each spatial channel divert a modulated digital reference signal from an artificial guide star off-chip for phase measurement. Science light from other wavelengths is coherently combined using on-chip beam combiners and outputted to a single waveguide. We described the role, design, fabrication and characterization of the photonic chip. This photonic phase control scheme can be applied in astronomical interferometry or optical satellite communications.

## 1. INTRODUCTION

### 1.1 Background

Astronomical interferometry uses multiple telescope apertures, or multiple subapertures, to create a high resolution image from the interference pattern between spatially separated receivers. In the former case of multiple telescopes, the imaging resolution can be extended to match that of a single telescope with the separation between the telescopes, a technique known as aperture synthesis[1]. When using a single large telescope, subapertures can be selectively opened to achieve diffraction limited imaging without sacrificing angular resolution for bright targets, a technique known as aperture masking[2]. In either case, the phase difference between light from separate apertures must be preserved, requiring adaptive optics and phase delay controls for multiple large telescopes, or only phase delay control for diffraction limited telescopes arrays.

The Very Large Telescope (VLT) is capable of performing such types of interferometry to achieve unprecedented angular resolution using its array of moveable telescope and the VLTI GRAVITY instrument[3]. Our implementation integrates the phase shifting and the beam combining onto a single chip, leveraging the phase correction speed advantages of integrated microheaters to perform rapid phase correction sufficient to correct atmospheric phase differences at well over 10 kHz, faster than fiber stretchers and highly scalable.

## 1.2 Overview

We present an integrated photonic phase shifter and beam combiner (PSBC) chip which can accept light from multiple telescopes or subapertures on a single telescope, and combine them coherently to produce visibilities for astronomical interferometry. Our method relies on the use of a satellite laser guide star[4,5] which can emit a bright digital signal modulation (DSM) signal in the form of a square wave laser pattern that is received by each aperture. This method is outlined in the proceedings by Herriot et al at this conference[6]. This paper describes the design, fabrication, and characterization of the photonic chip that would be used in the concept approach by Herriot. By modulating the laser at VHF(very high frequency) radio frequencies of ~100 MHz, the dynamic range of the phase measurement from each telescope is on the order of metres, as opposed to using the frequency of the carrier wavelength of light which is limited to ~1.5 microns in H-band before phase wrapping occurs[7]. Since the DSM signal passes through the same atmospheric column as the science light, it shares a spatial wavefront distortion and delay that is highly correlated as the science light. Some of the biggest challenges to long baseline optical interferometry are the large phase differences experienced by separate telescopes due to atmospheric turbulence, and the rate of change of such phase differences. Using an integrated photonic PSBC chip, these challenges can be mitigated as the dynamic range of the phase shifting on a chip can be in the hundreds of microns with a few hundred milliwatts[8], and the speed of integrated photonic phase shifters can exceed 100 kHz[9]. An illustration of the satellite guide star with a DSM laser signal is shown in Figure 1.

The light from each telescope is guided into a PSBC chip fabricated on the silicon-on-insulator (SOI) platform. We use established and new integrated photonics components designs to realize the PSBC operation on a single chip. We describe the simulated and characterized performance metrics of PSBC chip. The chip design, fabrication, characterization, microheater operation, and optical coupling are also discussed.

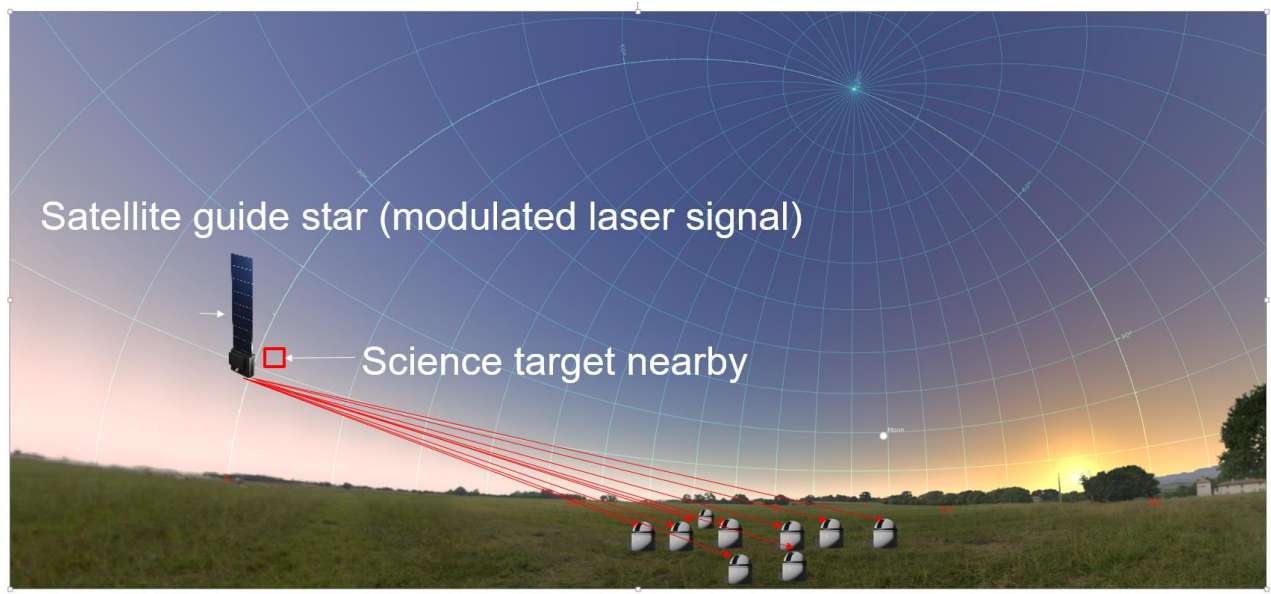

Figure 1. Illustration of the DSM signal from a satellite guide star and science light captured by multiple telescopes on the ground.

## 2. SIMULATION AND DESIGN

### 2.1 System operation

The PSBC chip operates by first coupling light from multiple receivers through a 1D fiber array. The light is through fed through separate phase delay lines in the form of long heated waveguide lengths folded into Archimedean spirals. The phase-corrected light is then spectrally filtered to extract a fraction of the DSM channel from the artificial laser guide star. The science light is combined coherently and coupled into a single waveguide. The DSM signal from the combined beam is then filtered again to provide a calibration signal. The filtered DSM signals from the multiple input channels are send to an array of fast photodetectors for phase measurement and control of the thermo-optic phase shifters. The science light output from the beam combiner is sent to a single photodiode for detection and generating the visibility fringes. A schematic of the system is shown in Figure 2.

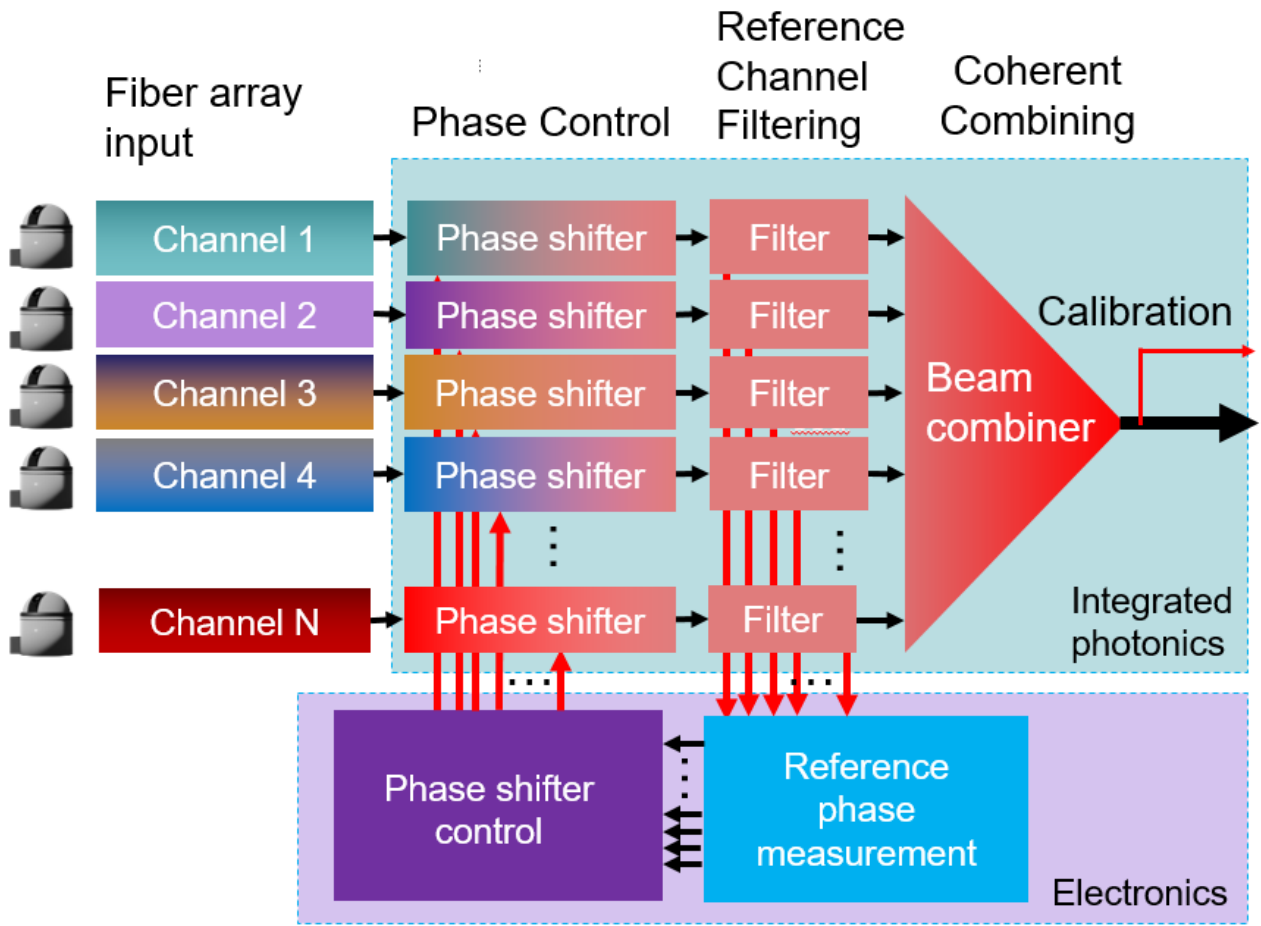

Figure 2. System schematic for the integrated photonic PSBC chip. The boxes in the green shaded region are operations performed on-chip. The boxes in the purple shaded region are electronic steps for phase measurement using photodetectors, and thermo-optic phase shifter control using power amplifiers.

## 2.2 Waveguides

Waveguides were designed to be strip waveguides with dimensions of 450 nm wide and 220 nm high to preserve single mode operation for TE polarized light. For this proof-of-concept, we have developed the PSBC chip for only TE polarization, but operation can be extended to both polarizations in parallel on the same chip. To accommodate all input light without polarization loss, a polarizer is required to split the light into two beam, with the TM polarized light would be injected into another waveguide array tailored to support to the TM polarization. The mode profile is shown in Figure 3. The effective index of was calculated to be 2.34, and group index is 4.37 of the waveguide mode at 1550 nm.

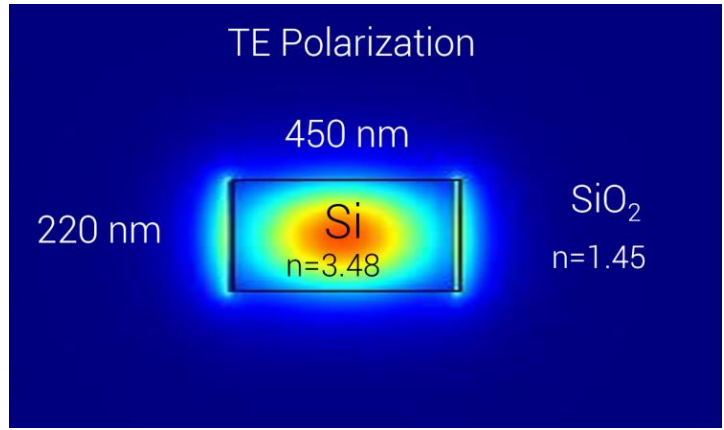

Figure 3. TE mode of the silicon nanophotonic waveguide.

## 2.3 Mode converters/edge couplers

The PSBC chip is designed to accept light from a fiber ribbon, which holds an array of single mode polarization maintaining lensed fibers. The light is coupled into the chip using mode converters to optimize the coupling efficiency between the lensed fibers tips spot and the waveguide mode. We use the high efficiency subwavelength grating edge couplers designs by Cheben et al. due to their high efficiency broadband operation[10]. The channels are physically separated on the chip by 127 µm, which is a standard fiber spacing in fiber ribbons. A scanning electron micrograph of the mode converter edge coupler is shown in Figure 4.

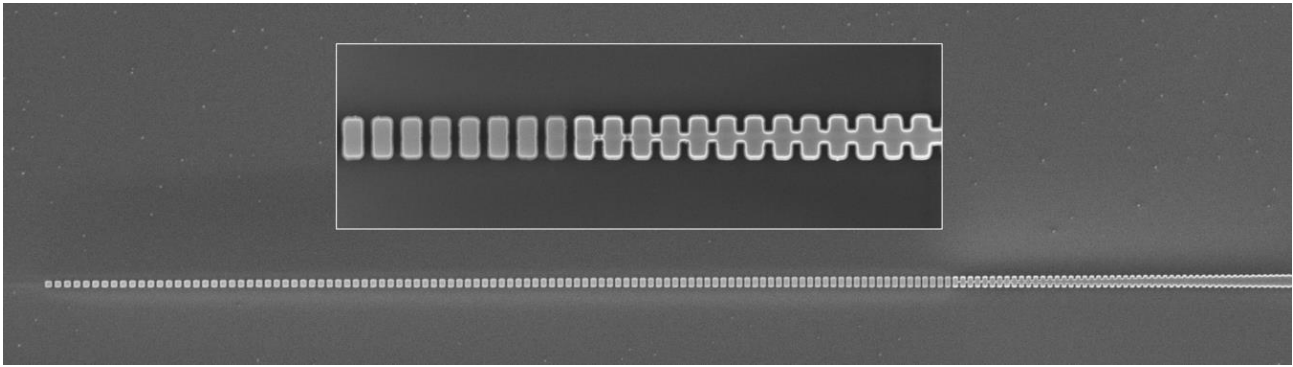

Figure 4. Scanning electron micrograph of the subwavelength edge coupling mode converters. The narrow end with discontinuous subwavelength segments faces a lensed fiber.

## 2.4 Beam combiners

Beam combiners were imagined in two configurations: star couplers, and hierarchical 2x1 combiners. The star couplers were designed to couple in light from multiple inputs into a single output using a free propagation region, commonly employed in pairs to form an arrayed waveguide gratings for telecommunication or spectrograph applications. The light from all waveguide "subapertures" on the left side of the star coupler is focused onto a single tapered receiver waveguide on the right. A simulated field profile, and geometry of the star coupler are shown in Figure 5a and 5b, respectively.

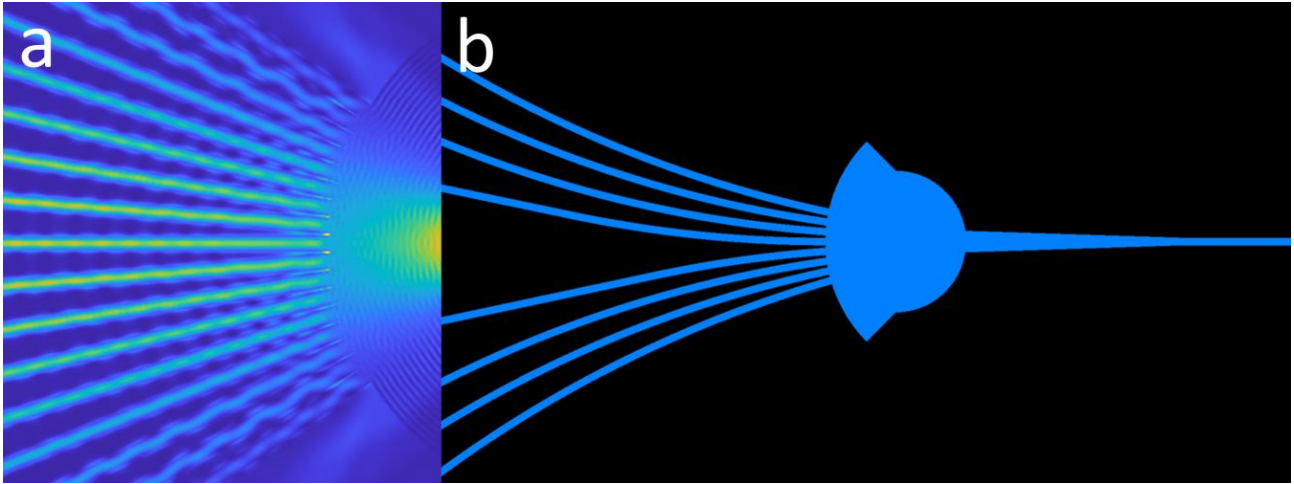

Figure 5. (a) Simulated $E^2$ field profile of the star coupler beam combiner, and (b) geometry of the fabricated star coupler.

The simulation in Figure 5a is performed with light being inputted from the right, leading to non-uniform output at the waveguides on the left. During actual operation, the light intensity from each waveguide will nominally be the same.

**2.5 Digital modulation channel filtering**

Our method relies on the use of an artificial guide star which produces a DSM at around 100 MHz. Light from this square wave modulation must be separated from the science light in order to provide the rapid, dynamic phase control without sacrificing the throughput of the science light. The wavelength of this channel can be chosen to be somewhere in the H-band. We have chosen a ring resonator-based spectral filter for its simplicity, low insertion loss, compactness and ease of thermal tuning. An optical micrograph of one of the ring resonator test structures is shown in Figure 6.

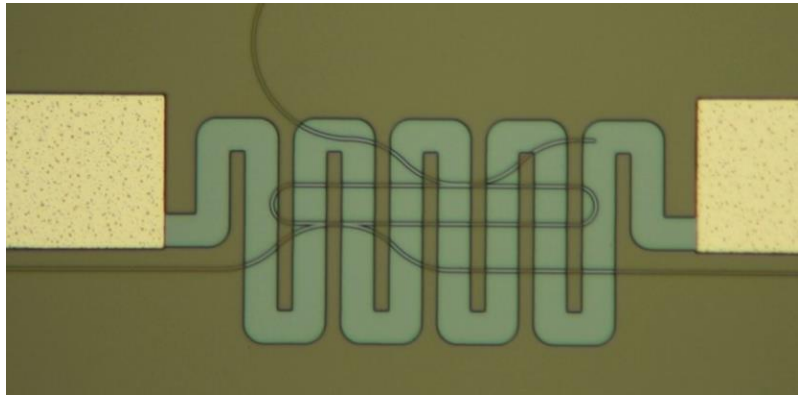

Figure 6. Optical micrograph of the fabricated ring resonator with patterned microheaters.

**2.6 Thermo-optic phase shifting**

Phase control between spatial channels are individually controlled using thermo-optic phase shifters in the form of serpentine microheaters. The temperature of the waveguides can be precisely controlled through an array of DC voltage sources. The resistance of the microheaters are designed to be approximately 50 ohms, with a trace width of 25 µm and length of approximately 400 µm. The temperature of the waveguides with 250 mW of heat dissipation per heater is calculated using the finite element thermal solver Lumerical HEAT. The cross-sectional thermal profile of the heaters with embedded waveguides below is shown in Figure 7.

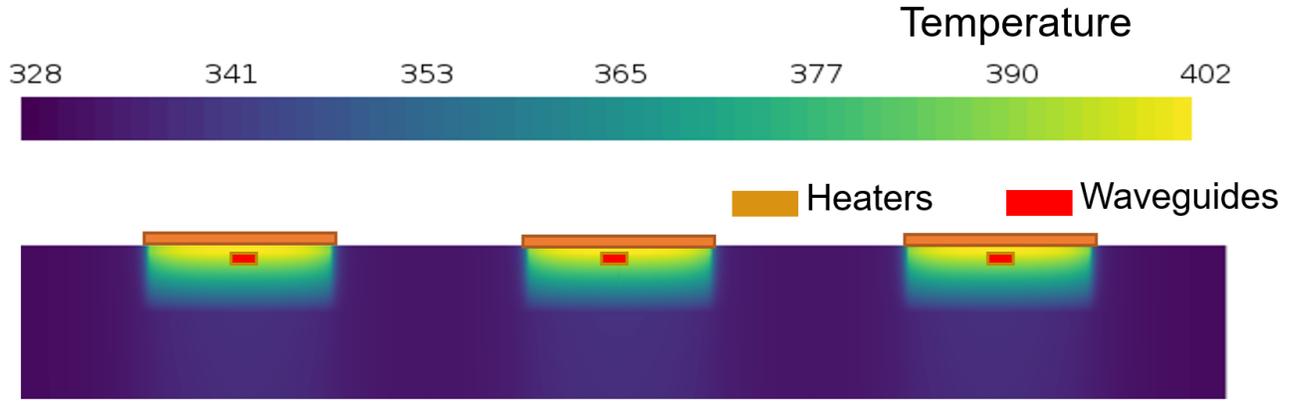

Figure 7. Simulated temperature profile through the chip across three adjacent heater spirals.

The temperature of the waveguides can reach 380 K, with the heaters themselves heated to just over 402 K. The magnitude of the phase shift can be approximately by:

$$\varphi = \left(\frac{2\pi}{\lambda}\right)\sigma_{TO} L(\Delta T) \quad (1)$$

where $\sigma_{TO}$ is the thermo-optic coefficient of the silicon waveguide mode, $\varphi$ is the phase shift magnitude in periods, $\lambda$ is the wavelength, $L$ is the length of the heated waveguide, $\Delta T$ is the temperature change. The thermo-optic coefficient of silicon is quite high at $1.86 \times 10^{-4}$ refractive index units (RIU) per degree, which is one of the main advantages of the SOI platform. We patterned phase delays lines with lengths ranging from ~2 mm to 20 mm. The number of periods per degree of heating is shown in Figure 8.

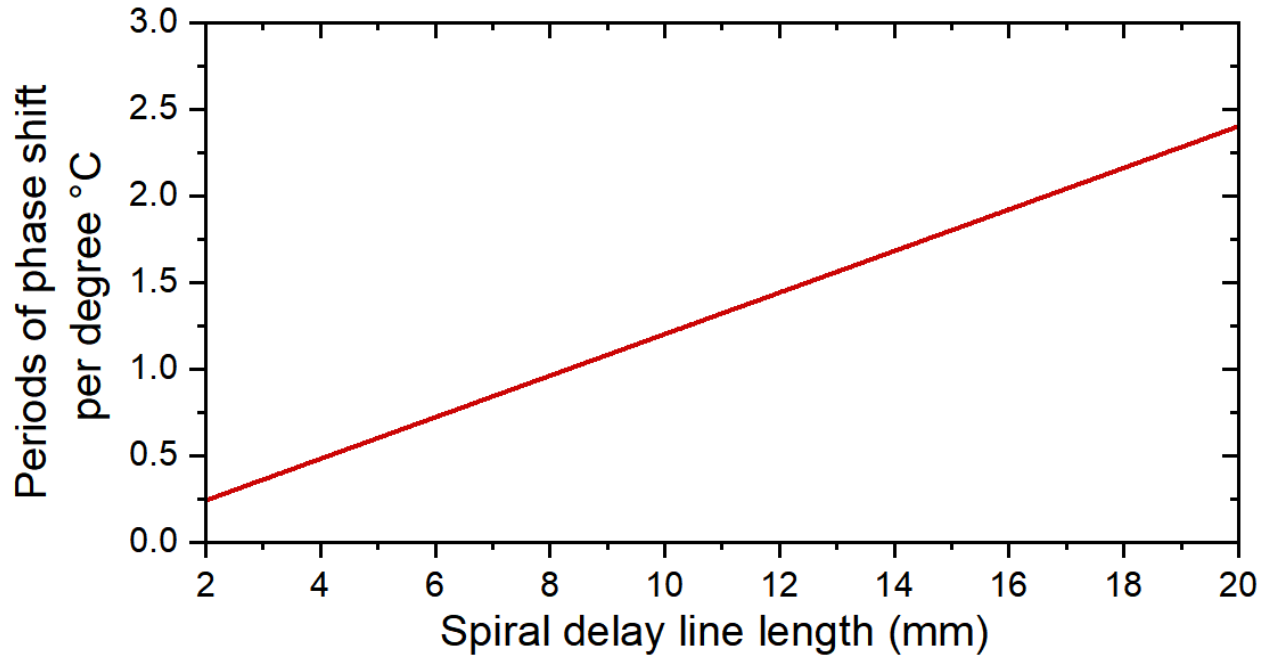

Figure 8. Simulated phase shift in periods per degree heating of the waveguides, as a function of the spiral delay line length.

The phase shift can be as high as 2.4 periods per degree for spiral delay lengths of 20 mm. This enables phase shift dynamic ranges of >180 µm with waveguide temperatures over 50 °C. With propagation losses on the order of 1 dB/cm, the trade-off in throughput must be considered as at least 2 dB (>37%) of loss would be expected for such a delay line length. This loss can be mitigated by designing for the silicon nitride platform, with the much reduced loss from high quality silicon nitride compensating for its reduced thermo-optic coefficient.

## 3. EXPERIMENTAL RESULTS

### 3.1 Fabrication

The chip was fabricated by Applied Nanotools Inc., a commercial silicon photonics chip foundry in Canada, using electron beam lithography. The buried silicon dioxide layer is 2 µm thick. The edges chips were dry etched to form facets for fiber coupling to the chip. The heaters are 200 nm thick TiW traces with contact pads made from a bilayer of 200 nm thick TiW + 500 nm aluminum. The 2.2 µm silicon dioxide cladding on top of the waveguides separates the heaters from the silicon waveguide. Eight total dies were fabricated, with each die supporting variations of PSBC devices and test structures. Devices with 1x1, 2x1, 4x1 and 8x1, and 32x1 channel PSBC devices were fabricated. One of the eight fabricated dies is shown in Figure 9, which has 1x1, 2x1, 8x1 devices.

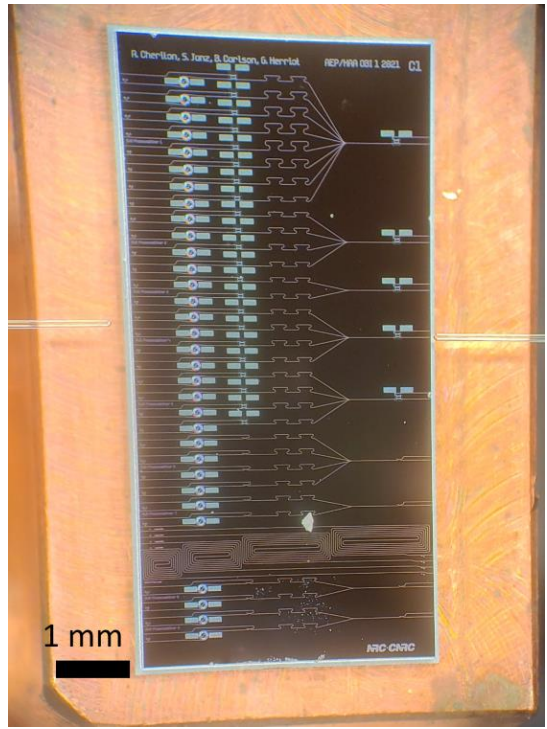

Figure 9. Photograph of one of the dies and lensed fibers on each side of the chip during characterization.

## 3.2 Characterization

The chips were characterized using a Keysight 81608A mainframe with a tunable laser and cooled InGaAs detector. The laser light polarization was rotated to couple into the TE mode of the waveguide. The chip was mounted on a temperature controlled stage with lensed fibers aligned to the waveguides on opposite sides of the chip. The propagation loss of the waveguide was determined using four loss structures with variable lengths, shown in Figure 10 (top). The slope of the insertion loss as a function of wavelength is shown in Figure 10 (bottom), which is approximately 0.9 dB/cm around 1550 nm.

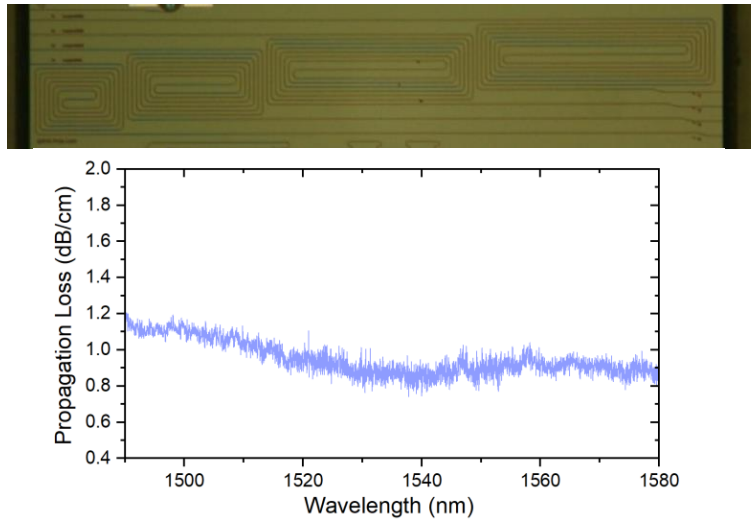

Figure 10. (Top) Optical micrograph of the waveguides used for loss estimation. (Bottom) Calculated propagation loss as a function of wavelength.

The TiW microheaters were connected to a Keithley 2400 Sourcemeter for sourcing and electrical characterization. Electrical contact to the chip was made using needle probes making contact to the exposed aluminum pad. Due to slight differences in the fabrication of ring resonators, there are usually differences between the transmission spectra of all ring resonators on photonic chips. To ensure the DSM signal is properly filtered by both ring resonators, both must be tuned to each have one of their resonances overlap with the designated DSM channel wavelength. Figure 11 shows that two ring resonators DSM filters (one before beam combining and one after beam combining) can be tuned into resonance with each other by applying current to one ring resonator microheater.

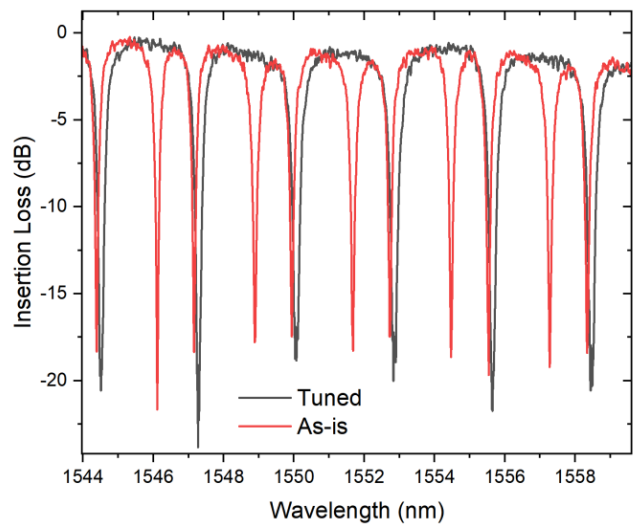

Figure 11. Normalized insertion loss of the two DSM ring resonator filters before (red) and after tuning into resonance (black).

The 8x1 star coupler beam combiner was characterized by guiding light backwards through the device, where the combiner acts as a beam divider, and measured at each of the output channels. This method removes any possible effect of decoherency and unequal intensity in the input channels. The throughput of the beam combiner was determined by comparing the insertion loss to 1x1 test devices (without the star combiner). Figure 12 shows the transmission spectrum of the beam combiner, with a throughput calculated to be approximately -2 dB (63%) without including any other components.

The ring resonators introduce losses of 14% of the science light due to the periodic nature of the filter, which can be further reduced by designing filters that are sharper and have larger free spectral range. These losses can be improved with sharper resonances and higher free spectral ranges.

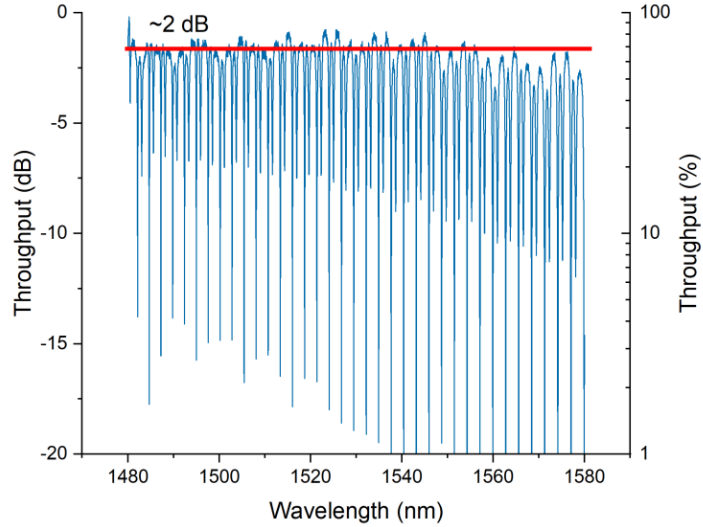

Figure 12. Estimated throughput of the 8x1 star coupler with two ring resonator DSM filters. The 2 dB throughput estimation is shown in red as a guide to the eye and does not include the loss from the ring resonances.

Without the beam combiner, the devices exhibit throughputs of approximately 17%, as shown in Figure 13. Including the star coupler losses, spiral delay line, DSM filters, and fiber-to-chip and chip-to-fiber coupling, the overall throughput of the 8x1 device is about 11% with phase delays spiral lengths of 2 mm. The throughput drop at longer wavelengths is a result of the increased loss with wavelength in the spiral delay lines and the lower efficiency of the star coupler away from 1550 nm. Total loss breakdown by component for a 100 nm bandwidth is estimated in Table 1 for our current chip, as well as an expected maximum throughput after design and fabrication optimization on the silicon nitride platform.

Table 1. Component and supplier for the demonstration system.

|  | Estimated throughput of our fabricated devices on SOI | Estimated maximum after optimization on silicon nitride |
|---|---|---|
| Edge coupling (output) | 63% | 93% |
| Edge coupling (output) | 63% | 93% |
| Propagation and phase delay spirals | 71% | 98% |
| Star coupler beam combiner | 63% | 89% |
| DSM Filter (phase) | 80% | 98% |
| DSM Filter (calibration) | 80% | 98% |
| **Total** | **11%** | **74%** |

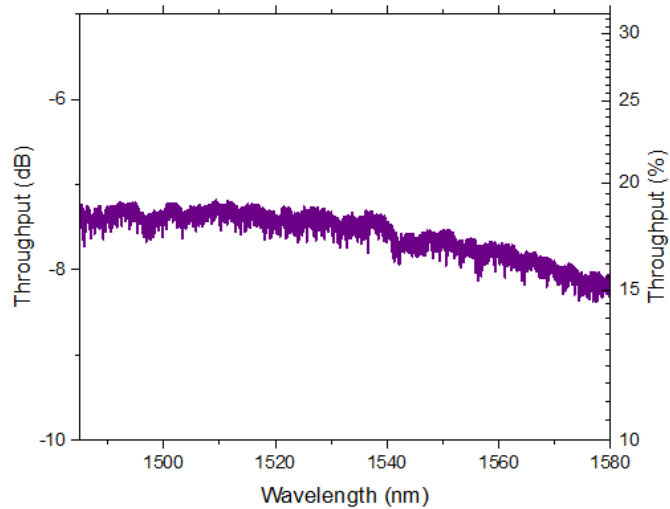

Figure 13. Throughput of the PSBC device without the beam combiner.

## 4. SUMMARY

In summary, we present our progress towards realizing a PSBC chip capable of rapid multichannel phase control and coherent beam combining for astronomical interferometry in the optical domain. We have designed PSBC devices to accept light from multiple optical fibers with differing phase and combine them coherently using mode converters, ring resonators, star couplers, and spiral delay lines with microheaters. We show the initial characterization of individual components of the PSBC devices. The 8x1 photonic chip has a throughput of about 11%. We show that the microheaters can be used to tune filters for a laser guide star signal into resonances to account for fabrication variations. Future work involves improving the throughput, designing for the silicon nitride platform, and demonstrating full operation of the chip on the optical bench with a higher number of input channels.

## REFERENCES

[1]    Brouw, W. N., "Aperture Synthesis," Image Processing Techniques in Astronomy, C. De Jager and H. Nieuwenhuijzen, Eds., 301–307, Springer Netherlands, Dordrecht (1975).
[2]    Tuthill, P. G., Monnier, J. D. and Danchi, W. C., "Aperture masking interferometry on the Keck I Telescope: new results from the diffraction limit," Interferometry in Optical Astronomy **4006**, 491–498, SPIE (2000).
[3]    Gravity Collaboration, Abuter, R., Accardo, M., Amorim, A., Anugu, N., Ávila, G., Azouaoui, N., Benisty, M., Berger, J. P., Blind, N., Bonnet, H., Bourget, P., Brandner, W., Brast, R., Buron, A., Burtscher, L., Cassaing, F., Chapron, F., Choquet, É., et al., "First light for GRAVITY: Phase referencing optical interferometry for the Very Large Telescope Interferometer," Astronomy and Astrophysics **602**, A94 (2017).
[4]    Marlow, W. A., Carlton, A. K., Yoon, H., Clark, J. R., Haughwout, C. A., Cahoy, K. L., Males, J. R., Close, L. M. and Morzinski, K. M., "Laser-Guide-Star Satellite for Ground-Based Adaptive Optics Imaging of Geosynchronous Satellites," Journal of Spacecraft and Rockets **54**(3), 621–639 (2017).
[5]    Thompson, W. and Marois, C., "Extremely bright orbital guide beacons for extremely large telescopes," presented at AO4ELT 2019: Proceedings 6th Adaptive Optics for Extremely Large Telescopes, 9 June 2019, Collection / Collection : NRC Publications Archive / Archives des publications du CNRC, 159215, AO4ELT6.
[6]    Herriot et al., "High contrast and high resolution sensing and correction of atmospheric turbulence without WFSs and DMs using a digital signal modulated satellite beacon and integrated photonics devices," SPIE Astronomical Telescopes + Instrumentation 2022 (This conference), Paper 12185-10 (2022).

[7] Halir, R., Vivien, L., Le Roux, X., Xu, D.-X. and Cheben, P., "Direct and Sensitive Phase Readout for Integrated Waveguide Sensors," IEEE Photonics J. **5**(4), 6800906–6800906 (2013).
[8] Qiu, H., Liu, Y., Luan, C., Kong, D., Guan, X., Ding, Y. and Hu, H., "Energy-efficient thermo-optic silicon phase shifter with well-balanced overall performance," Opt. Lett. **45**(17), 4806 (2020).
[9] Tong, W., Wei, Y., Zhou, H., Dong, J. and Zhang, X., "The Design of a Low-Loss, Fast-Response, Metal Thermo-Optic Phase Shifter Based on Coupled-Mode Theory," Photonics **9**(7), 447 (2022).
[10] Cheben, P., Schmid, J. H., Wang, S., Xu, D.-X., Vachon, M., Janz, S., Lapointe, J., Painchaud, Y. and Picard, M.-J., "Broadband polarization independent nanophotonic coupler for silicon waveguides with ultra-high efficiency," Opt. Express **23**(17), 22553 (2015).